\documentclass[twocolumn,showpacs,preprintnumbers,amsmath,amssymb]{revtex4-1}

\usepackage{graphicx}
\usepackage{dcolumn}
\usepackage{bm}

\begin{document}

\preprint{COMS6998 - Network Theory - HW3}

\title{Motif Analysis in the Amazon Product Co-Purchasing Network }

\author{Abhishek Srivastava}
 \affiliation{Computer Science Department, Columbia University}
 \email{aas2234@columbia.edu}

\date{\today}

\begin{abstract}
Online stores like Amazon and Ebay are growing by the day. Fewer people go to departmental stores as opposed to the convenience of 
purchasing from stores online. These stores may employ a number of techniques to advertise and recommend the appropriate product to
the appropriate buyer profile. This article evaluates various 3-node and 4-node motifs occurring in such networks. Community structures are evaluated too.These results may provide interesting insights into user behavior and a better understanding of marketing
techniques.
\end{abstract}

\maketitle

\section{\label{sec:level1} Introduction}
According to a recent study, the number of people choosing to shop at a physical departmental store is on a decline. More and more people today are using the convenience of online shopping \cite{nytimes}. The growth of sites like Amazon and Ebay is evidence of this emerging trend. Stores also tend to encourage buyers with discounts and e-coupons for shopping online since it means that they do not have to stock up goods at their brick and mortar stores but may ship them as and when required from the warehouse directly to the buyer. This is now a well proven business model for lowering operational costs and maximizing profits.

As a result of this trend, stores online are able to profile buyers based on their preferences. Stores may use such information to recommend products to other buyers showing a similar profile or may cleverly advertise the products that a buyer is more likely to buy. In either case, the advertising engine would need to analyze large datasets of information containing past records of buyers, the products they bought and the timing of their purchase. Leskovec et all \cite {Leskovec} used such a dataset from a large online retailer to study the effectiveness of viral marketing techniques. They researched the effectiveness of personal recommendations between buyers and modeled these based on a simple stochastic model. Viral marketing is a proven technique as demonstrated by the growth of email services like Yahoo and Hotmail who based their marketing on this technique. Each email sent out from their email services contained a footer advertising the service. Simple, and very effective.

Referral marketing is surely the future of web advertising. Free services online tend to lose their value if they are easily available to even those who don't need them. Google used a referral technique to launch their GMail mail service by allowing users only by invitation. This requires some effort from the potential user of the free service tool to actively seek out users to try and get a referral. It also gives current users a value perception of the free product they are using because it is not widely available to everybody. 

This article will aim to look at the Amazon product co-purchasing network, on both a micro as well as a macro scale. Understanding the interactions at a localized and micro scale may help in isolating the noise and point to interesting trends that can be extrapolated to the larger scale. Motifs have been used extensively to study interactions in biological networks, especially in bacteria like the E-Coli. The article shows how motifs can be used in a product co-purchasing network to gain insights into relations between products. Both 3-node and 4-node motifs have been analyzed.

Community structures have long been used in network analysis and the article by Leskovec \cite{Leskovec} identified communities and the products that these communities preferred. In this article, we suggest using community detection methods to make the motif-based analysis more accurate and pertain to a select list of products. 

Section II describes the source and the structure of the dataset we will use for analysis. Section III concentrates on the motifs discovered in the dataset, their significance and their frequency of occurrence. Section IV discusses algorithms for detecting community structures in the dataset for identifying trends between products frequently co-purchased together.

\section{\label{sec:level2} Nature and Description of the Dataset}

The dataset for this article was obtained from the publicly available Stanford Network Analysis Platform \cite {snap}. The description reads as follows : ``Network was collected by crawling Amazon website. It is based on Customers Who Bought This Item Also Bought feature of the Amazon website. If a product \texttt{i} is frequently co-purchased with product \texttt{j}, the graph contains a directed edge from \texttt{i} to \texttt{j}.'' Along with the edge-list representation of co-purchased products, there is also metadata present for each product summarized in \texttt{Table I}. 

\begin{table}
\caption{\label{tab:table1} Sample Amazon Product Metadata}
\begin{tabular}{|l|c|}
\hline
\bf{Property} & \bf{Value}\\ \hline
{\bf Id}  &   1\\ \hline
{\bf ASIN} & 0827229534\\ \hline
{\bf Title } & Patterns of Preaching: A Sermon Sampler\\ \hline
{\bf   Group } & Book\\ \hline
{\bf   SalesRank } & 396585\\ \hline
{\bf   Similar Products } & 5  0804215715  156101074X  \\ & 0687023955  0687074231  082721619X\\ \hline
{\bf   Categories } & 2
   |Books[283155]|Subjects[1000]|\\ & Religion \& Spirituality[22]| \\ & Christianity[12290]
|Clergy[12360]|\\ & sPreaching[12368]\\ \hline
  {\bf Reviews } & Total: 2  downloaded: 2  avg rating: 5\\ &
    2000-7-28  cutomer: A2JW67OY8U6HHK \\ & rating: 5  votes:  10  helpful:   9 \\ &
    2003-12-14  cutomer: A2VE83MZF98ITY \\ & rating: 5  votes:   6  helpful:   5\\ \hline

\end{tabular}
\end{table}

This article's experiments were carried out on a subset of the original dataset to restrict the computational resources required to carry them out. The portion of the dataset which will be used for analysis in this article was collected on March 02 2003. Some statistics of this dataset are presented in \texttt{Table II}. As depicted by the statistics, the network is quite dense with a clustering coefficient of about 0.42. The fraction of nodes belonging to the largest strongly connected component is very high - 0.922. This implies that the products sampled are very closely related with each other.

\begin{table}
\caption{\label{tab:table2} Dataset Statistics}
\begin{tabular}{|l|c|}
\hline
\bf{Property} & \bf{Value}\\ \hline
{\bf Nodes } &	262111 \\ \hline
{\bf Edges} &	 1234877\\ \hline
{\bf Nodes in largest WCC } & 262111 (1.000)\\ \hline
{\bf Edges in largest WCC } &	1234877 (1.000)\\ \hline
{\bf Nodes in largest SCC } &	241761 (0.922)\\ \hline
{\bf Edges in largest SCC } &	1131217 (0.916)\\ \hline
{\bf Average clustering coefficient }&	0.4240\\ \hline
{\bf Number of triangles } &	717719\\ \hline
{\bf Fraction of closed triangles } &	0.2361\\ \hline
{\bf Diameter (longest shortest path)} &	29\\ \hline
{\bf 90-percentile effective diameter } &	11\\ \hline

\end{tabular}
\end{table}

\section{\label{sec:level3} Motifs : Description and Significance}

\subsection{\label{sec:level1} Introduction to Motifs}

Motifs can be described as recurring, significant patterns of interconnections present in a network. These motifs have been found to occur more frequently than in random graphs; a number of biological networks are composed solely of a large number of such motifs.Each type of network seems to display its own set of characteristic motifs (ecological networks have different motifs than gene regulation networks, etc). Milo et all analyzed such network motifs in their famous article \cite {milo} where they showed that motifs formed the basic building blocks of a number of biological networks and they also formulated methods to detect the commonly occurring motifs in a network. 

\subsection{\label{sec:level2} Market Segmentation }
Buyers online, as anywhere else, are usually classified into profiles that are indicative of their past buying trends. Market segmentation is a marketing term referring to the aggregating of prospective buyers into groups (segments) that have common needs and will respond similarly to a marketing action. Market segmentation enables companies to target different categories of consumers who perceive the full value of certain products and services differently from one another. Generally three criteria can be used to identify different market segments:

1) Homogeneity (common needs within segment)

2) Distinction (unique from other groups)

3) Reaction (similar response to market)

\subsection{\label{sec:level3} Analysis Functions for Motifs}

In this section, we will define a product purchasability function for a 3-node motif that will indicate the probability of product being purchased in the context of the 3-node motif using in-degree ${V^{i}_{in}}$ at vertex \texttt{i} and ${|E_{motif}|}$, the total count of edges in the motif : 

\begin{equation}
f(P_{i}) = \frac{{V^{i}_{in}}}{|E_{motif}|}
\end{equation}
 
The product purchasability function is not defined for a vertex in the motif with in-degree zero. For all other vertices, this function will return the fraction of edges coming into the product as opposed to the total number of edges. To rate each motif, we will then declare a Motif Rank function that will indicate the fraction of nodes that have a finite, positive product purchasability function. Contrary to what the name of this function implies, it does not identify the importance or significance of a motif. Instead, this function looks at the fraction of nodes (products) whose purchasability may be defined and thus provide a larger actionable product-set.
\begin{equation}
 \textnormal{Motif Rank} = \frac{\textnormal{Number of Nodes with Positive $f(P_{i})$}}{\textnormal{Total Nodes in Motif}} 
\end{equation}

 It must be noted that the Product Purchasability and the Motif Rank functions are comparable only within the particular class of k-node motifs (where k is usually 3 or 4) for which the functions were defined. We will now look at some of the most commonly occurring motifs discovered in the product co-purchasing network in the next sections.

\subsection{\label{sec:level4} Analysis of 3-Node Motifs in Network}

\begin{figure}[h!]
  \caption{Motif ID : 1}
  \centering
    \includegraphics[scale=0.2]{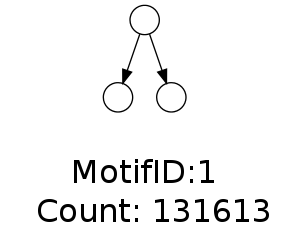}
  \label{motif000}
\end{figure}

The statistics of 3-node motifs present in the network studied in this article have been summarized in \texttt{Table III}. As seen in biological networks, the probability of occurrence of some motifs is much higher (by an order of magnitude) than others. This is indicative of a structural similarity present in the network. Figure \ref {motif000} shows a 3-node motif with a frequency count of 131613 and MotifID 1. Motif IDs were assigned to the commonly occurring 3-node and 4-node motifs by the original article on motifs by Milo et all \cite{milo}.

\begin{table}
\caption{\label{tab:table3} }
\begin{tabular}{|c|c|c|c|}
\hline

{\bf MotifId} & {\bf Nodes} & {\bf Edges} & {\bf Count}\\ \hline

1 &	3 &	2 &	131613\\ \hline
2 &	3 &	2 &	78578\\ \hline
3 &	3 &	3 &	104071\\ \hline
4 &	3 &	2 &	217566\\ \hline
5 &	3 &	3 &	16090\\ \hline
6 &	3 &	4 &	20400\\ \hline
7 &	3 &	4 &	32962\\ \hline
8 &	3 &	4 &	3685\\ \hline
9 &	3 &	3 &	2\\ \hline
10 &	3 &	3 &	135904\\ \hline
11 &	3 &	4 &	21579\\ \hline
12 &	3 &	5 &	28397\\ \hline
13 &	3 &	6 &	19319\\ \hline

\end{tabular}
\end{table}

 MotifID 1 shown in Figure \ref {motif000} is a representation of the case where when a customer bought a product, he / she went ahead and bought either or both of the other 2 products. The $f(P_{i})$ of nodes in this motif is 0.5 for the nodes at the bottom and undefined for the node at the top. We assume that the time instant at which the product represented by the top node is the present instant. The customer may then purchase both the other products, either of the 2 products or none of them.  The motif rank can then 
 be computed as 0.66.

\begin{figure}[h!]
  \caption{Motif ID : 3}
  \centering
    \includegraphics[scale=0.2]{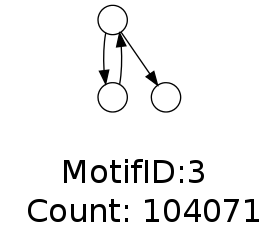}
  \label{motif002}
\end{figure}

 MotifID 3 shown in Figure \ref {motif002} is a interesting motif with a relatively high frequency of occurrence. There is a reciprocating relation between 2 of the nodes. This is a strongly connected component indicating the closeness similarity between the nodes. The $f(P_{i})$ of the nodes in this motif are all 0.33. The Motif Rank function for this motif is found to be 1.  

\begin{figure}[h!]
  \caption{Motif ID : 4}
  \centering
    \includegraphics[scale=0.2]{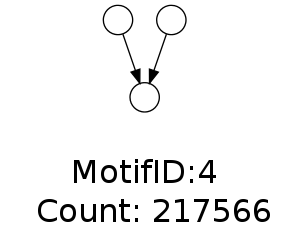}
  \label{motif003}
\end{figure}

 MotifID 4 shown in Figure \ref {motif003} has the highest frequency of occurrence among all the 3-node motifs. The next highest frequency of occurrence is about 40\% lower than the frequency of this motif. This a converging motif - consumers who bought largely unrelated products (the 2 nodes at the top) also bought the product represented by the node at the bottom. The unrelated nature of the 2 product nodes at the top stands out in this motif. The high frequency of this motif also points to the fact that there is a converging factor in the network implying that a large number of customers end up purchasing a small subset of products. Identifying this small subset of products can be used by the online store for improving services and reducing operational costs by smartly stocking this identified subset of products. The Motif Rank of this motif is 0.33 due to the presence of just one node with a positive $f(P_{i})$ function. 

\begin{figure}[h!]
  \caption{Motif ID : 10}
  \centering
    \includegraphics[scale=0.2]{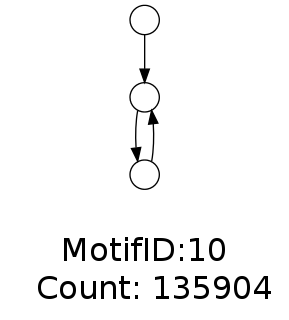}
  \label{motif009}
\end{figure}

MotifID 10 shown in Figure \ref{motif009} contains a strongly connected component showing a close correlation between the products. The $f(P_{i})$ of the node at the top is undefined. The node in the centre has a $f(P_{i})$  of 0.66 and the node at the bottom has one of 0.33. Customers buying the product represented by the node at the top may buy the one in the middle too. The argument can be extended to the product at the bottom too. With increasing edge separation between product nodes, the probability that the customer will buy the product falls exponentially. This is due to the fact that most customers place budgets and prioritize the items to purchase depending on the ones they need the most.
   
\subsection{\label{sec:level5} Analysis of 4-Node Motifs in Network}

\begin{figure}[h!]
  \caption{4-Node Motif Frequency Distribution}
  \centering
    \includegraphics[scale=0.45]{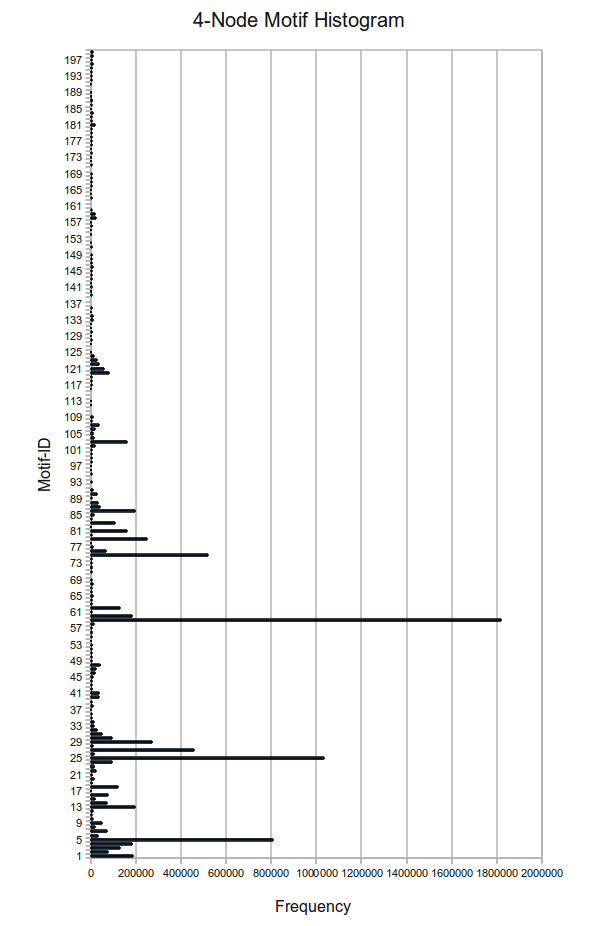}
  \label{4-motif-hist}
\end{figure}

In this section, we used a motif searching algorithm to detect all 4-node motifs in the network and obtained their frequency of occurrences. Motif IDs range from 1 to 200, encompassing all the possible permutations of 4-node motifs. Figure \ref{4-motif-hist} shows the frequency distribution of the 4-node motifs. The three tallest bars belong to Motif ID 59, 26 and 5.

\begin{figure}[h!]
  \caption{Motif ID : 5}
  \centering
    \includegraphics[scale=0.2]{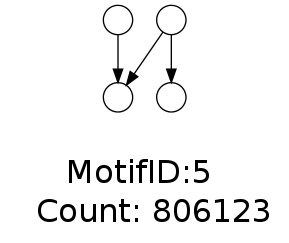}
  \label{4-motif004}
\end{figure}

\begin{figure}[h!]
  \caption{Motif ID : 25}
  \centering
    \includegraphics[scale=0.2]{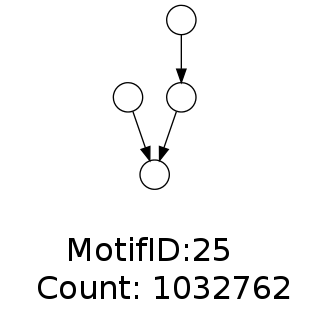}
  \label{4-motif024}
\end{figure}

\begin{figure}[h!]
  \caption{Motif ID : 59}
  \centering
    \includegraphics[scale=0.2]{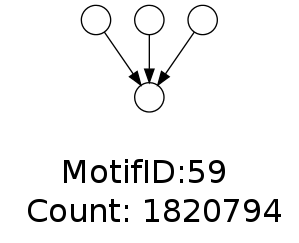}
  \label{4-motif058}
\end{figure}

Figures \ref{4-motif004}, \ref{4-motif024} and \ref{4-motif058} show the 3 most frequently occurring 4-node motifs in the product co-purchasing network.The Motif IDs 25 and 59 share a characteristic similar to that observed in the most frequently occurring 3-node motif with MotifID 4. They are both converging motifs indicating the high probability of a customer buying the product to which the edges in the motif converge. The interesting resemblance to the analysis done for 3-node motifs is that the 4-node motif with Motif ID in Figure \ref{4-motif058} also has the highest frequency of occurrence among all 4-node motifs.

Figure \ref{4-motif004} is the third most frequently occurring 4-node motif in the network. The $f(P_{i})$ of the 2 nodes at the bottom are 0.66 and 0.33. This motif is neither converging nor diverging. But there is a trend in the motif to move toward the bottom node with a higher $f(P_{i})$, indicating a higher probability of the product represented by this node to be purchased.

Its interesting to observe that the Motif ID 25 shown in Figure \ref{4-motif024} appears in second place, statistically. This can be identified as a converging motif with all nodes tracing a path to the bottom node. Again, the $f(P_{i})$ function of the bottom node is the highest among the nodes in the motif. This motif may also represent a similarity in tastes from customers belonging to different schools of thought or having different backgrounds. 

Motif ID 59 is truly indicative of the convergence trend mentioned previously. It is also statistically the most significant 4-node motif, just like its similar looking counterpart among the 3-node motifs. 

\subsection{\label{sec:level6} Extrapolating the Micro to the Macro }

Motifs, by themselves do not describe the large scale structure of the network. To infer the properties of the larger picture, we will have to aggregate the perspective provided by the individual motif analysis. The suggested method to obtain the most mileage would be to break the original network into a number of communities using the Girvan-Newman algorithm \cite{gir-new} or the faster Clauset, Neuman, Moore algorithm \cite{clauset}. Once this is done, each of these communities represents a singular class of objects (nodes). If our network is large and varied, we can continue this process till we reach our required coarseness of sub-categorization. These resulting networks can then be subjected to motif analysis as as carried out in the previous sections. 

In case of the product co-purchasing network used in this article, such granularity will enable us in studying the dynamics of particular products more accurately. Since buyer profiles are built with progressive purchases which may not belong to the same community of products, we cannot extrapolate the findings obtained from these community-centric entities to build customer profiles. What we may use them for is building product purchasing trends. We can monitor the motif statistics over time to gather useful information about trends. For example, if the frequency of occurrence of converging motifs increases, we deduce that a larger set of customers are looking to buy a fewer, smaller subset of products. Having adjusted the coarseness of the community sub-categorization, we can make a good guess as to what these products are and whether, as online retailers,  we need to make any operational changes to match the demand.

\section{\label{sec:level4} Community Dectection for Motif Analysis}

Social networks are a product of the contexts that bring people together. The
context can be a shared interest in a particular topic or kind of a book. Some-
times there are circumstances, such as a specific job or religious affiliation, that
make people more likely to be interested in the same type of book or DVD. 

Community discovery algorithms presented in \cite {gir-new} and \cite{clauset} provide mechanisms for community detection in complex systems. The Girvan-Neuman algorithm \cite {gir-new} focuses on identifying edges that are the weakest in the network, the ones that are the least central . These would inadvertently be edges that are connecting communities. Then communities are detected by progressively removing such edges from the original graph. The modularity maximization method \cite {clauset} detects communities by searching over possible divisions of a network for one or more that have particularly high modularity. Since exhaustive search over all possible divisions is usually intractable, practical algorithms are based on approximate optimization methods such as greedy algorithms, simulated annealing, or spectral optimization, with different approaches offering different balances between speed and accuracy.The usefulness of modularity optimization is however questionable: on the one hand, it has been shown that modularity optimization often fails to detect clusters smaller than some scale, depending on the size of the network (resolution limit); on the other hand the landscape of modularity values is characterized by a huge degeneracy of partitions with high modularity, close to the absolute maximum, which may be very different from each other.The choice of algorithms to use for community detection may be made at the behest of the investigator who knows the nature of the network being studied.

\section{\label{sec:level5} Conclusions and Future Work }

The article defined some functions for analyzing the occurrence and nature of motifs present in a network. Motifs, by themselves are not very much helpful in feature extraction,as observed, and we must combine such analysis with other methods like community detection. Nevertheless, we have shown that motifs can be used to visualize trends, extract interesting large scale features and on the whole, be useful in predicting user behavior in a network such as a product co-purchasing network. 

With reference to a co-purchasing network, we were able to show that the following traits may be extracted using motifs : 

(1) Prediction of product demand 

(2) Purchasing Trends in product categories

(3) Relations between different products

The extraction of these trends is vital from the perspective of online retailers hoping to cash in on the growing number of customers choosing to shop online. The one striking trend that this article found was that there is a large presence of converging motifs, indicating that most customers tend to buy a select few products. This is not a contradiction to the ``long tail'' phenomenon. On the contrary, it can be shown that the presence of the long tail is a large scale phenomenon and would not make itself visible at a micro sampling of the network. As \cite {Leskovec} has shown, the long tail is the future of online retailing; selling less of more is more pronounced in online retailing than in physical retailing.

In the future, one can look more deeply into the interactions between motifs and into the cascading effects of these motifs in the context of the network neighborhood. Various trends may be identified and used effectively for improving marketing and advertising of products by the management of a retailing organization. 
\bibliographystyle{plain}

\begin{thebibliography}{99}

\bibitem{Leskovec}
  Leskovec, Adamic, Huberman
  \emph{The Dynamics of Viral Marketing}
  ACM Transactions on the Web, Vol. 1, No. 1, Article 5, May 2007

\bibitem{snap}
    SNAP
    \emph{Stanford Network Analysis Platform} 
    http://snap.stanford.edu

\bibitem{milo}
  Milo,Shen-Orr,Itzkovitz,Kashtan,Chklovskii,Alon
  \emph{Network Motifs: Simple Building Blocks of Complex Networks}
  Nature, Vol 298, 25 October 2002 

\bibitem{nytimes}
  Robust Sales for Holiday Weekend
  \emph{http://www.nytimes.com/2010/11/29/business/economy/29shop.html}
  New York Times, 28 November 2010

\bibitem{gir-new}
  Girvan-Newman algorithm (Girvan M. and Newman M. E. J., 
  \emph{Community structure in social and biological networks}
   Proc. Natl. Acad. Sci. USA 99, 7821<96>7826 (2002))

\bibitem{clauset}
  A. Clauset, M.E.J. Newman, C. Moore
  \emph{Finding Large community in networks}
   2004




\end{thebibliography}

\end{document}